# A pragmatic approach to regulating AI agents

*Philipp Hacker[1] and Matthias Holweg[2]*

*Abstract*

*The current advancement in and deployment of agentic AI systems has created a set of key challenges for the legal frameworks that govern their use. We cover two central components: first, the regulatory classification of agents under the EU AI Act, and second, the legal status and validity of autonomous actions within the established framework of EU contract law. We argue that the unique capacity of agents to autonomously reason, plan, and execute tasks across disparate external systems necessitates a fundamental shift in oversight toward the orchestration layer, where multi-agent interactions introduce novel risks of misalignment. While agents generally utilise general-purpose AI models, we posit that their structural complexity and cross-system permeability require them to be regulated as "AI systems" with distinct obligations under the AI Act. Consequently, our proposals highlight the need for robust accountability mechanisms to manage this heightened autonomy. On the contractual side, we advocate for a "traffic light" system of staggered task authorization based on operational risk and the creation of a statutory list of non-delegable legal acts. By implementing these measures, we provide a pragmatic pathway to ensure that the increasing autonomy of AI agents remains firmly anchored in human accountability and existing legal standards.*

**Working paper**

Version 2, April 15, 2026

Comments and suggestions welcome!

---

[1] European New School of Digital Studies, European University Viadrina. E: hacker@europa-uni.de
[2] Saïd Business School, University of Oxford. E: matthias.holweg@sbs.ox.ac.uk

# 1   Introduction

The advent of agentic AI systems signals another transformative wave of AI adoption, with projections suggesting the deployment of billions of agents within the next decade. Industry forecasts indicate that by the end of 2026, 40% of enterprise applications will feature task-specific AI agents—a significant surge from less than 5% in early 2025 (Gartner, 2025). At the same time, a stark governance gap is emerging: while 75% of organizations intend to deploy agents by 2028, only 21% currently possess a mature governance framework to oversee them (Deloitte, 2026). This chasm is amplified by persistent ambiguity how agentic AI systems fit into existing regulatory frameworks, like the AI Act, and what legal status decisions made by agentic AI systems carry under EU contract law. This paper addresses the current misalignment between rapid technological advancement and the lag of a robust legal framework to regulate and govern this powerful technology.

While the concept of an "autonomous agent" is an established paradigm in computer science, it has long remained an unfulfilled promise due to computational limitations (Wooldridge, 1997; 2009). The emergence of the transformer architecture (Vaswani et al. 2017) and the Large Language Models (LLMs) derived from this architecture, has catalysed a renaissance in workflow automation, yet these systems introduce novel risks stemming from their heightened autonomy and cross-system permeability. These concerns mirror the warnings of Geoffrey Hinton and Yoshua Bengio, who emphasize the dangers of systems that learn through opaque processes, the potential for autonomous resource acquisition, and the risk of self-replication (Bengio et al., 2023; 2026). Agentic AI theoretically amplifies these risks; by design, agents possess greater agency than standalone systems, frequently interacting with external environments to exchange instructions and information. Consequently, the threats of model misalignment and "alignment faking" are significantly more pronounced in agentic architectures than in traditional AI (Ji et al., 2023; Jiang et al., 2025; Koorndijk, 2025).

Against this backdrop, this paper investigates a central question: How can we design a legal framework capable of regulating a class of AI characterized by a distinctly higher degree of autonomy? We analyse this autonomy through two critical lenses: first, we consider regulatory compliance by evaluating whether the EU AI Act is "fit for purpose" regarding the specific risks of agentic behaviour; second, we consider the contractual status of determining how autonomous actions are governed under EU contract law, specifically regarding the validity and limits of decisions made by agents within their deployed contexts.

We begin by elucidating the key differences between traditional AI, General-Purpose AI (GPAI), and AI agents. We then elaborate on the legal consequences of these distinctions under the AI Act, before assessing the implications of autonomous agency for established contract law. We conclude with a suite of policy proposals aimed at regulating the use of agentic AI and providing a clear legal context for their autonomous actions. Certain areas of law, such as data protection (Nannini et al., 2026), consumer (Busch, 2025; Kolt, 2025), and copyright law (Henderson et al., 2023; Quintais, 2025) remain beyond the scope of this article.

## 2   The classification of AI agents: How are they different?

Although commonly referred to as a single construct of "artificial intelligence" or simply "AI", in practice the way AI is implemented encompasses many different forms. This heterogeneity is both a function of technological advancements – for example by replacing classical machine learning algorithms with deep learning or neural networks – as well as a function of different contexts in which AI is deployed, as certain algorithms perform much better in some contexts than others – for example, transformer architectures and their ability to process natural language. The key theoretical point here is commonly referred to as the "no free lunch theorem" (Wolpert and Macready, 1997), which states that no single algorithm performs well in all circumstances. In other words, each distinct application or context will require a different algorithm.

Different algorithms and applications require different regulatory approaches, as the risks inherent in their deployment differ significantly. In the context of the AI Act this heterogeneity is reflected in several ways: the general definition of different risk levels, the specification of prohibited and high-risk categories of AI deployment, as well as the provisions for general-purpose AI models. In this context it is therefore essential to delineate AI "agents" from the other AI systems.

Technically, agents differ significantly from standalone AI systems in several important ways. First and foremost, in terms of structural complexity, agentic AI combines both stochastic and deterministic components. GPAI models, which function probabilistically, serve as reasoning engines and language interfaces that translate prompts into the "next best action" through chains of thought (CoT). The proposed action is, in effect, stochastic as it is produced by an LLM, and subject to the same technical constraints. This process allows agents to reason and act autonomously in response to complex (i.e, non-standardised) user inputs. Agents then connect to other systems (via protocols like MCP) or other agents (via protocols like A2A), to form multi-agent systems. This part is generally deterministic, as formats for API calls require standard communications.

The second key distinction is the remit for agency, or scope to act. Agency is the key defining feature of agentic AI, consisting primarily of reasoning and action. Reasoning refers to the ability to interpret user prompts in order to make a decision based on the "next best action" identified. It is important to emphasise here that the reasoning part of an agent, in the large majority of cases to date, relies on iterative chain-of-thought (CoT) "reasoning" by a large language model. While other sources of knowledge, like knowledge graphs, retrieval augmented generation, mixture of experts (MoE) and links to external systems like deterministic calculators may be embedded in the agent's architecture, in essence, the fundamental and well-documented limits to reasoning still apply (Felin and Holweg, 2024; Mirazadeh et al., 2024; Shojaee et al., 2025).

Action, by contrast, refers to the autonomous execution of these decisions, often through interfaces with external systems. In many cases, agentic AI systems may also communicate and collaborate with other agents through multi-agent orchestration, forming complex AI-enabled workflows that operate with limited human oversight (Dang et al., 2025). Referring back to Bengio et al. (2026), it is this access to other systems (so called "privileges") that pose potential danger. In practice this includes an agent's ability to execute financial transactions, to interfere with critical infrastructures like electricity grids, or to govern access to information.

The third, and most obvious, distinguishing feature of agentic AI is the level of autonomy granted to them. In traditional systems, humans remain in the decision-making loop ("human in the loop"), whereas agentic AI operates, or at least may operate, independently with only "human on the loop" oversight. In such circumstances, the human no longer is involved ("in the loop") of each individual AI-based decision but only supervises the operation of the AI system as a whole.

Once provided with an initial prompt or objective, agents can act on their own initiative, executing actions identified by the system without requiring further human direction. This marks a key distinction that in practice is often blurred: AI *tools* or *workflows* execute a set of steps only once prompted by a user; AI *agents* execute a set of steps autonomously without specific permission or instigation. An example of the former is a search-and-retrieval workflow like a customGPT that is triggered by the user when needed; an example of the latter would be an agent that monitors a customer account and autonomously executes actions like sending marketing communications, or suggested actions to the customer.

In terms of technical composition, agents use ReAct (reasoning-action) frameworks (cf. Yao et al., 2022) in most cases: it starts with a reasoning part that analyses the input prompt and uses a CoT logic to identify the next best action (recommender logic). The agent then seeks to define the steps for action (execution), for which it connects to other systems and agents (via MCP or A2A protocols, amongst others) to execute that action. Reasoning uses a reasoning LLM, CoT iterative prompting. So "action" is an LLM output, which generally is probabilistic. Acting refers to the execution, which often features interfaces to other systems and agents and generally is deterministic. Hence, as mentioned, agents tend to feature both probabilistic and deterministic features.

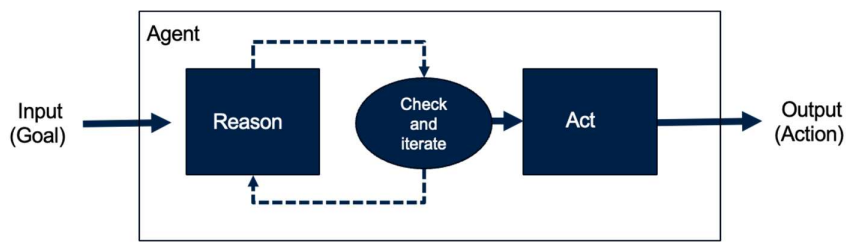

**Figure 1: Iterative Reasoning-Acting (ReAct) Framework of an AI Agent. Based on Yao et al. (2022)**

Beyond the definition of a single AI agent, it is important to state that in practice agents will not be operating in isolation, but in concert with other agents (Aranda and Sugimoto, 2026). There are several reasons for this, mostly of an economic nature: it is easier (i.e. less costly) to build, test and deploy agents with a smaller scope. Also, the ability to reuse agents in a different context or setting increases if the agent is less specific, thus providing economies of scale.

The general task of coordinating multiple agents is commonly referred to as multi-agent orchestration. Orchestration encompasses a range of complex tasks, and includes for example:

1. **Task decomposition**, i.e. breaking a complex "mega-prompt" into a logical sequence of smaller, manageable sub-tasks.
2. **Agent routing**, i.e. dynamically selecting and assigning the most qualified specialized agent (e.g., Coder vs. Analyst) to each sub-task.
3. **State management**, i.e. maintaining a "shared memory" so agents can pass context, files, and findings to one another without losing data.

4. **Conflict resolution**, i.e. managing "debates" between agents when their outputs or goals contradict each other.
5. **Quality control**, i.e. implementing a "reviewer" agent to verify the accuracy of the "worker" agent's output before it proceeds.
6. **Ensuring Human-in-the-Loop feedback**, i.e. inserting strategic pauses for human approval during high-stakes or low-confidence steps
7. **Cost optimization**, i.e. routing tasks to cheaper or faster models when high-tier reasoning is not required to save on compute.
8. **Error recovery**, i.e. detecting "looping" behaviour or tool failures and re-triggering tasks with a new strategy.

Recent academic work and practice also suggests that LLMs can function not only as end-task generators but also as controllers that orchestrate multiple agents, route inputs to appropriate models and coordinate modular planning components within hybrid systems combining general-purpose and specialized models (Shnitzer et al., 2023; Webb et al., 2025; Yang et al., 2025).

As deployment expands across more tasks and more reasoning-intensive workflows, token consumption and inference costs can rise sharply; empirical and theoretical work on inference economics and routing shows that routing many requests through large commercial models can materially increase cost; this makes hybrid routing strategies attractive for controlling spending and preserving quality (Erdil, 2025; Bergemann et al., 2025; Shnitzer et al., 2023; Ding et al., 2024). In this sense, coupling agents with smaller language models or specialized models may reduce both financial and environmental costs, because smaller models generally require less compute at inference time and can lower energy consumption while preserving acceptable task quality in agentic systems (Ngo-Ho et al., 2026; Belcak et al., 2025; Jeanquartier et al., 2026).

To conclude, AI agents are generally built with a LLM-based front-end to reason but will vary greatly in terms of their structure to execute any action. Agents generally are deployed in workflows comprising other agents, while again LLMs will be used to orchestrate these agents.

## 3  Agentic autonomy and the AI Act: Classification and relevant provisions

To understand the implications of regulation for agentic AI, it is useful to consider how the EU AI Act applies in its current form, but also how agents are embedded into the current framework of contract law.

### 3.1  AI systems and AI models

The EU AI Act (Regulation (EU) 2024/1689) draws a fundamental regulatory boundary between "AI systems" and "general-purpose AI models" (GPAI models). Each of them is subjected to distinct obligations enforced by different authorities. Famously, while a definition of an AI system exists in the AI Act (Article 3(1)), the term "AI model" remains undefined in the Act. In our view, as we shall show in greater detail below, an AI model is a trained mathematical artifact – a necessary but insufficient component of an AI system, which additionally requires a user interface, data pipelines, deployment infrastructure, and/or a defined operational context (see also Fernández-Llorca et al., 2025, 881).

Hence, for the Claude AI system, Opus 4.6 is one model used; for the original ChatGPT system, GPT 3.5 was the model powering the system.

This distinction reflects both a genuine technical difference recognized across computer science (below, 3.1.2) and a pragmatic regulatory adaptation driven by the rise of generative AI (Boine & Rolnick, 2024; Veale & Quintais, 2025; Zenner, 2025): the Act's original risk-based framework, built around systems with intended purposes, could not accommodate general-purpose models that lack predetermined use cases. The result is a regulatory architecture in which system-level obligations follow a fourfold risk classification (prohibited, high-risk, limited, minimal) and are generally enforced by national authorities. By contrast, model-level obligations follow a two-tier structure under a dedicated Chapter V (transparency and copyright rules for all general-purpose AI models; systemic risk provisions for general-purpose AI models with systemic risk), enforced centrally by the EU AI Office rather than by national authorities (Novelli et al., 2024).

Article 3(1) defines an "AI system" as "a machine-based system that is designed to operate with varying levels of autonomy and that may exhibit adaptiveness after deployment, and that, for explicit or implicit objectives, infers, from the input it receives, how to generate outputs such as predictions, content, recommendations, or decisions that can influence physical or virtual environments."

This definition, aligned with the OECD's November 2023 updated definition, centres on the system's capability to infer, which Recital 12 clarifies "transcends basic data processing by enabling learning, reasoning or modelling" (European Commission, 2025b). Hence, an AI system is a complete operational entity that interacts with an environment, receives inputs, and generates outputs that influence that environment. Importantly, in the product safety logic under which the AI Act was originally conceived and finally enacted, such a system is endowed with a specific intended purpose, which in turn drives the risk classification (Fernández-Llorca et al., 2025, 877).

Raw AI models do not have such a purpose. While the definition of an AI model is missing, the AI Act does contain a term of a general-purpose AI model (Veale & Quintais, 2025). Article 3(63) defines it as "an AI model, including where such an AI model is trained with a large amount of data using self-supervision at scale, that displays significant generality and is capable of competently performing a wide range of distinct tasks regardless of the way the model is placed on the market and that can be integrated into a variety of downstream systems or applications".

Two features of Article 3(63) are striking. First, the definition uses the term "AI model" without the Act ever independently defining it. This makes an AI model an autonomous concept of EU law, but simultaneously seems to treat the concept as a commonly understood technical term (below, 3.1.2). Second, the definition emphasizes the model's generality and its capacity to be integrated into a variety of downstream systems. This positions a (GPAI) model as an upstream component that feeds into downstream AI systems (European Commission, 2025a, 2025c).

Article 3(66) provides a bridge concept for the GPAI world, defining "general-purpose AI system" as "an AI system which is based on a general-purpose AI model and which has the capability to serve a variety of purposes, both for direct use as well as for integration in other AI systems." This definition again makes explicit that when a GPAI model is wrapped in additional components, such as an interface, and endowed with a deployment context or an operational purpose, it becomes a GPAI system (see also Fernández-Llorca et al., 2025; Gutierrez et al., 2023). This, in turn, is a subcategory of an AI system under Article 3(1).

Several additional definitions reinforce the model-system boundary. Article 3(3) defines "provider" to cover both AI systems and GPAI models but reveals an asymmetry: a provider may place an AI system or GPAI model "on the market" (e.g., by offering it for sale) but only an AI system can be put into service (e.g., by being put into use).[3] This demonstrates, again, that a model, unlike a system, is not itself operationally deployable. Correspondingly, Article 3(4) defines "deployer" exclusively in terms of AI systems ("a natural or legal person [...] using an AI system under its authority"), not models. One deploys a system; one does not deploy a model. Capturing further value chain scenarios, Article 3(68) defines "downstream provider" as "a provider of an AI system [...] which integrates an AI model". The technical integration provides the definitional hinge connecting the model layer to the system layer.

The following table captures the regulatory logic embedded in these definitions:

| Regulatory concept | AI system | GPAI model |
|---|---|---|
| Can be "placed on the market" (Art. 3(9)) | Yes | Yes |
| Can be "put into service" (Art. 3(11)) | Yes | **No** |
| Has a "deployer" (Art. 3(4)) | Yes | **No** |
| Has a "provider" (Art. 3(3)) | Yes | Yes |
| Subject to fourfold, purpose-based risk classification | Yes (Arts. 4-7, 50) | No (different two-tier classification: horizontal + systemic risk, Art. 51-33) |
| Requires conformity assessment / CE marking | Yes (high-risk, Art. 43) | **No** |

**Table 1: Definition and embedded regulatory logic.**

The recitals, while not strictly legally binding, are often used by regulators and courts to interpret the binding articles and illuminate the legislator's reasoning. Recital 97 is the most important one in our context:

> "Although AI models are essential components of AI systems, they do not constitute AI systems on their own. AI models require the addition of further components, such as for example a user interface, to become AI systems. AI models are typically integrated into and form part of AI systems."

This last sentence clarifies the distinction: the model is a component, the system is the whole. Recital 97 also addresses the overlap scenario where a provider integrates its own model into its own system, specifying that "the obligations in this Regulation for models should continue to apply in addition to those for AI systems". With this, it establishes the key principle of *cumulative* obligations across regulatory layers.

Recital 110 explains why models merit separate regulation despite not being systems: GPAI models "could pose systemic risks", including "negative effects in relation to major accidents, disruptions of critical sectors and serious consequences to public health and safety [...] the dissemination of illegal, false, or discriminatory content." Systemic risks "increase with model capabilities and model reach, can arise along the entire lifecycle of the model, and are influenced by conditions of misuse, model reliability, model fairness and model security." The risks are model-level properties that propagate through any downstream system built on the model; it is economically more efficient and computationally more effective to address them once at the source than in a multitude of cases downstream (Hacker et al., 2025).

When a GPAI model is integrated into a high-risk AI system, obligations stack: the upstream model provider complies with Chapter V (GPAI model provider rules), while the downstream system provider complies with Chapter III (high-risk provider rules), and Article 53(1)(b) requires the model provider to supply sufficient information for the system provider to fulfil its risk-management, data-governance, and transparency obligations. Article 25(4) specifies that the parties should formalize this information-sharing in a written agreement. This layered architecture reflects the recognition that model providers may see their products integrated into a range of downstream systems and must therefore enable, but cannot substitute for, system-level compliance (Hacker et al., 2023).

What remains unaddressed, though, is how specific additions to the model that do not yet reach the point of a fully deployable system should be treated: as still part of the model or already part of the system layer. For example, retrieval-augmented generation (RAG) components are close to the model itself. Nevertheless, we would qualify them as system components and not part of the model (Fernández-Llorca et al., 2025; Hacker & Holweg, 2026). A RAG component does not alter the mathematical properties (e.g., weights and biases) of the model itself. Furthermore, otherwise, adding or changing a RAG might easily convert the downstream deployer into the provider of a GPAI model, by analogy to Article 25(1)(b) AI Act (Hacker & Holweg, 2026). However, the Act was not intended to shift the compliance burden for Chapter V duties onto downstream modifying entities so quickly (Hacker & Holweg, 2026; Zenner, 2025). Concerning AI agents, we shall return to this delineation when discussing agentic scaffolding below (3.2).

Since the AI Act does not provide clear guidance on what a model is, we turn to computer science for inspiration. The drafters of the AI Act clearly chose a technical term; hence, a technical analysis can illuminate the interpretation of that wording. In fact, the AI Act's model-system distinction maps onto well-established technical concepts. In computer science, a model is a mathematical or statistical artifact – typically a parameterized function learned from data – while a system is the complete operational assembly that deploys one or more models alongside surrounding infrastructure.

In Goodfellow, Bengio, and Courville, a model is characterized through the recurring notation:

$$y = f(x; \vartheta)$$

This parameterized function that maps inputs (x) to outputs (y), with the parameter θ adjusted via a learning algorithm (Goodfellow et al., 2016, 6, 110, 116, 163). In this sense, it is a purely mathematical construct with no deployment context, user interface, or operational infrastructure.

Bommasani et al. (2021) at the Stanford Center for Research on Foundation Models coined the term "foundation model" and explicitly characterized these models as having a "critically central yet incomplete character" (Bommasani et al., 2021, 1). Similarly, Zaharia et al. (2024) from Berkeley AI Research provide one of the clearest contemporary definitions: "We define a compound AI System as a system that tackles AI tasks using multiple interacting components, including multiple calls to models, retrievers, or external tools. In contrast, an AI Model is simply a statistical model, e.g., a Transformer that predicts the next token in text."

A widely cited paper by Sculley et al. provides a vivid illustration: their Figure 1 shows the ML model code as a tiny central box surrounded by vast infrastructure for data collection, data verification, feature extraction, configuration management, process management, serving infrastructure, and monitoring (Sculley et al., 2015). Again, the model alone cannot function in the real world; the system is what operates.

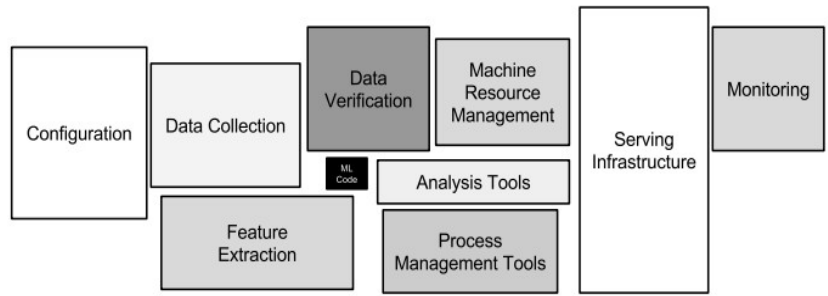

**Figure 2: Composition of LLMs, Source: Sculley et al. (2025:4)**

Concerning agents more specifically, the foundational AI textbook by Russell and Norvig frames AI through the concept of intelligent agents that perceive their environment via sensors and act upon it through actuators. In this framework, a trained model (e.g., a neural network for classification) is called an agent function. It constitutes merely one component of the agent's internal architecture; the model is implemented by an agent program, and the agent/system additionally encompasses perception, reasoning, decision-making, and action (Russell & Norvig, 2022, 55-56).

International standards bodies have formalized this hierarchy found in the computer science literature. For example, ISO/IEC 22989:2022 formally distinguishes the model ("A machine learning model is a mathematical construct that generates an inference, or prediction, based on input data or information ", Definition 5.11.7) from the AI system ("engineered system that generates outputs… for human-defined objectives", Definition 3.1.4).

In summary, the EU AI Act's distinction between AI systems and GPAI models is both technically sound and regulatorily necessary, but it introduces structural tensions that scholars continue to debate. The core insight is simple: a model is a learned mathematical artifact; a system is the operational whole that deploys it. This maps cleanly onto computer science definitions and explanations (Sculley et al., 2015; Zaharia et al., 2024), international standards (ISO/IEC 22989), and the Act's own Recital 97. The regulatory consequence is the standard two-track framework: a fourfold risk-based classification for systems (prohibited, high-risk, limited, minimal), a twofold layer for models (horizontal-plus-systemic-risk obligations), with cumulative obligations where the two tracks intersect in a value chain.

The deeper challenge is that the Act was designed as product safety regulation for systems with intended purposes, and the GPAI model provisions were grafted on late to address a category of technology that defies purpose-based risk classification. The model-system boundary, rather clean in principle, will face growing pressure as AI development trends toward compound and orchestrated systems (Belova et al., 2026; Webb et al., 2025; Zaharia et al., 2024) that blur the line between what is "just a model" and what constitutes an operational system. AI agents are one phenomenon that muddies the waters further, as we now discuss.

### 3.2    *Regulation of agents as GPAI "models" or GPAI "systems"?*

A threshold question under the AI Act is whether AI agents should be classified as GPAI models, as AI or GPAI systems (Sections 3.2 and 3.3), or as yet another category (Section 3.5). We address each possibility in turn, including value chain considerations (Section 3.4).

At the outset, it is important to note that agents are typically not GPAI models. As set out above (Sections 2 and 3.1), AI agents are not mere mathematical artefacts. They are systems built on top of foundation models and connected to them through scaffolding architecture. An agent has a concrete instantiation – it features an interface, it is deployed in a specific environment, and it interacts with external tools and data sources. Therefore, it cannot be reduced to the underlying model weights and parameters that constitute a GPAI model within the meaning of Article 3(63) of the AI Act.

Agents do, however, qualify as AI systems (Oueslati & Staes-Polet, 2025; see also Nannini et al., 2026). An AI agent – understood as a software entity that autonomously plans, reasons, uses tools, and acts upon its environment to pursue goals (Kasirzadeh & Gabriel, 2025; Russell & Norvig, 2022; Sumers et al., 2024; see also Section 2) – squarely satisfies the definition of "AI system" in Article 3(1). It is a machine-based system that operates with "varying [i.e., sufficient] levels of autonomy," that infers from inputs how to generate outputs, and whose outputs "influence physical or virtual environments." Notably, actions or task execution are valid outputs under the AI Act, too (but see Moreau & Henning, 2026): the examples listed in Art. 3(1) ("predictions, content, recommendations, or decisions") are not exhaustive, and the actions rely on prior numeric or textual output by the underlying model. The autonomy element in the AI Act, despite conflating automation with autonomy (Hacker, 2024), appears almost tailor-made for agents: Recital 12 clarifies that AI system autonomy means "that they have some degree of independence of actions from human involvement and of capabilities to operate without human intervention."

Beyond simple AI systems, however, agents may additionally qualify as *GPAI* systems. Because agents are AI systems, they can also meet the definition of a general-purpose AI system under Article 3(66) if the underlying GPAI model enables them to tackle a sufficiently broad range of tasks (Oueslati & Staes-Polet, 2025, 21). Recall that Article 3(66) defines a "general-purpose AI system" as "an AI system which is based on a general-purpose AI model and which has the capability to serve a variety of purposes, both for direct use as well as for integration in other AI systems." Notably, the statutory criterion is capability, not actual use: even if a deployer restricts an agent to a narrow set of tasks, the agent qualifies as a GPAI system as long as it retains the capability to serve a variety of purposes.

This reading aligns with the Commission's Guidelines on the scope of obligations for providers of general-purpose AI models under the AI Act, which provide an indicative criterion for a model to be considered a GPAI model: its training compute exceeds $10^{23}$ FLOPs and it can generate language – whether in the form of text or audio – or produce text-to-image or text-to-video outputs (European Commission, 2025c, para. 17). The Commission further clarifies that the modality of "text" includes "code" (European Commission, 2025c, fn. 2). An agent built on such a model and capable of, for instance, general-purpose code generation would therefore satisfy both the model-level and the system-level thresholds. Similarly, an orchestration-level agent, built on a frontier large language model, routing tasks to test specific agents, supervising them and collecting results would likely cross the compute threshold and display sufficient generality. Such agents would qualify as a GPAI system, however; the underlying model – but not the agent itself – would be considered a GPAI model (see also Hacker & Holweg, 2026; Oueslati & Staes-Polet, 2025, 21).

The practical significance of this classification as a *GPAI* system remains limited, however. The distinction between mere AI systems and GPAI systems is, in regulatory terms, largely a theoretical one (see also Noller & Rappenglück, 2026, 17). The AI Act does not impose materially different obligations on GPAI systems as compared to other AI systems at the system level: the additional obligations under Chapter V attach to GPAI models and their providers, not to the system classification as such; and the high-risk rules apply only if the intended purpose specifically encompasses high-risk use cases the general regulatory architecture of the AI Act speak against qualifying GPAI systems as high-risk per se, only on the grounds that they could, theoretically, be used in high-risk contexts. This is true of many AI systems; hence, in our view, the intended purpose must specifically be geared toward one or several high-risk use cases (see also (Schwartmann & Zenner, 2025; Oueslati & Staes-Polet, 2025, 21). From a systematic perspective, a general purpose is not enough: it can only trigger GPAI model provider rules, not high-risk system rules. This is also evident from the formulation of Article 25(1)(c) AI Act, which stipulates that a GPAI system, which has not been classified as high-risk, becomes high-risk if it is used for a high-risk purpose. This rule would be obsolete if any GPAI system would automatically, irrespective of its use, be a high-risk system.

A harder question is whether certain general-purpose agentic frameworks could or should still be qualified as AI models – or even as GPAI models – rather than as systems, or whether they necessitate an altogether different regulatory category. We investigate this question below in Section 3.5, once we have gained a better understanding of their regulation as AI systems.

### 3.3    Regulation of agents as AI systems

The risk classification of an agent as a system depends on its deployment context. An AI agent that autonomously manages job applications could fall under Annex III, point 4 (employment, workers management) and thus qualify as high-risk. If the AI agent is intended to book tables at a restaurant or flights, it would typically qualify as limited-risk as it interacts with humans, but outside of high-risk categories. An AI agent that autonomously manipulates human behaviour through subliminal techniques could even trigger the prohibition in Article 5.

This context-dependency is precisely the product-safety logic that the Act inherited from the New Legislative Framework: risk is assessed at the system-in-use level, not at the technology level. As scholars have observed, the AI Act assigns risk to systems based on their intended purpose (Fernández-Llorca et al., 2025; Hacker, 2024; Veale & Borgesius, 2021), a feature that works well for traditional vertical applications but sits uneasily with general-purpose agents whose intended purpose is, by design, open-ended. Nevertheless, under the current AI Act, agents can be categorized along the four risk levels for AI systems depending on the specific deployment context.

**Transparency obligations under Article 50**

Article 50 of the AI Act applies to AI agents, and it does so with particular force where agents qualify as GPAI systems. The provision establishes a layered transparency regime that imposes obligations on both providers and deployers.

Under Article 50(1), providers must ensure that AI systems designed to interact directly with natural persons are marked in a way that informs those persons that they are interacting with an AI system – unless this is obvious from the circumstances and context of use. For AI agents, this obligation is of central practical relevance. Agents that autonomously communicate with users – whether through text, voice, or other modalities – must be labelled as artificial interlocutors. The more sophisticated and human-like the agent's conversational abilities, the more pressing the need for clear disclosure becomes.

However, not all interactions with AI agents need to be labelled as such – only direct interactions with humans. For example, if an agent books a ticket via an automated online reservation system, such an identification would be unnecessary as there is no human on the other end. If the agent then fails and writes an email to the restaurant, its AI basis again need not be disclosed. The output is directed towards humans, but the humans do not directly interact with the system (cf. Martini, 2026, para. 57-58). Only when the agent serves as a chatbot, interacting with humans in real time who engage with the system itself, the provider needs to ensure that it discloses its AI-based status. Note that this concerns the system provider; this will generally be the entity wrapping the model into an agentic scaffolding (downstream provider, see above).

Article 50(2) further requires providers of AI systems – which, as discussed, includes GPAI systems – that generate synthetic audio, image, video, or text content to ensure that the outputs are marked in a machine-readable format and are detectable as artificially generated or manipulated. For agents that produce such content autonomously, providers bear the responsibility to embed appropriate technical markers at the system level (e.g., watermarking).

Deployers, in turn, must under Article 50(4) inform natural persons that they are exposed to AI-generated or AI-manipulated audio, image or video content if this amounts to a deepfake. In contrast to the watermarking provision, this type of deepfake labelling must be salient and clearly visible. While most people think of manipulated images of humans, the notion of a deepfake under the AI Act extends much further. According to Article 3(60), it comprises an "AI-generated or manipulated image, audio or video content that resembles existing persons, objects, places, entities or events and would falsely appear to a person to be authentic or truthful". Hence, importantly, agents used in marketing or publishing contexts are covered if their non-textual output seems authentic, but is not. For example, Coca-Cola produced a Christmas advertisement on the basis of 70,000 AI generated videos (Landymore, 2025). When the output is used in the EU (Article 3(1)(c)), for example by broadcasting the advertisement in EU television, such marketing campaigns may need to be labelled as AI-generated, depending on the degree of artificial authenticity.

Text is treated differently under Article 50(4) AI Act. It only needs to be marked as AI generated if it is published with the purpose of informing the public on matters of public interest, and if it has not undergone a process of human review or editorial control or editorial responsibility is lacking. For most textual publications on matters of public interest, some kind of editorial control will typically be installed by the PR department of the releasing entity. However, agents could change this. If indeed an agent is used to directly post textual output online, which relates to matters of public interest, the output must be identified as AI-generated or manipulated. For example, this could be the case if companies let agents collect market or news overviews overnight and publish them early in the morning in a newsletter to clients or the general public. If there is no human editorial control between output and publication, labeling is necessary. Given the still high rate of hallucinations in LLMs and even specialized and curated AI systems (Chelli et al., 2024; Ji et al., 2023; Magesh et al., 2025; Xu et al., 2024; Zhao et al., 2024), this legal obligation seems fully justified.

Again, these obligations apply with equal force to agents that operate as GPAI systems. An agent built on a GPAI model and capable of autonomous content generation across a variety of domains must comply with the full suite of Article 50 requirements – irrespective of whether the deployer restricts the agent to a narrow use case or closely monitors the content. The transparency obligations attach to the system's capabilities and its mode of interaction, not to the deployer's particular configuration.

**Prohibitions under Article 5**

The prohibitions laid down in Article 5(1) of the AI Act apply to AI agents, understood as AI systems, without qualification. The intended purpose of the system does not pose a conceptual obstacle. A more interesting question arises where an agent does not have a designated prohibited purpose but is nonetheless used in a prohibited manner. Unlike the high-risk classification rules – which depend primarily on the system's intended purpose – the prohibition regime captures systems that were designed with a general or even a limited purpose and that are subsequently deployed in prohibited contexts, even if the provider did not envision such use.

The prohibitions in Article 5(1) can be grouped into three categories according to the nexus they require between the system and the prohibited conduct. In the first category, general capability suffices. For certain prohibitions, it is enough that the system can generally exhibit the prohibited behaviour, even if neither the provider nor the deployer intended this result. This category covers the placing on the market, the putting into service, or the use of an AI system that deploys subliminal techniques beyond a person's consciousness (Article 5(1)(a)); the placing on the market, the putting into service, or the use of an AI system that exploits any of the vulnerabilities of a natural person or a specific group of persons (Article 5(1)(b)); and the placing on the market, the putting into service, or the use of AI systems for the evaluation or classification of natural persons or groups of persons over a certain period of time based on their social behaviour – so-called social scoring (Article 5(1)(c)).

For the second category, a purpose-specific placement or use is required. For other prohibitions, the statute requires that the system be put on the market or into service precisely for the incriminated purpose. This category includes predictive policing – that is, the placing on the market, the putting into service for this specific purpose, or the use of an AI system for risk assessments of natural persons in order to assess or predict the risk of a natural person committing a criminal offence (Article 5(1)(d)). It also encompasses the placing on the market, the putting into service for this specific purpose, or the use of biometric categorization systems that categorize natural persons individually on the basis of their biometric data to deduce or infer their race, political opinions, trade union membership, religious or philosophical beliefs, sex life, or sexual orientation (Article 5(1)(e)); the creation or expansion of facial recognition data sets via Internet scraping or CCTV footage (Article 5(1)(f)); and emotion recognition systems in the workplace or education (Article 5(1)(g)).

In the third category, use alone is decisive. Concerning real-time remote biometric identification for law enforcement purposes, only the use – not the placing on the market or the putting into service – triggers the prohibition: Article 5(1)(h) prohibits "the use of 'real-time' remote biometric identification systems in publicly accessible spaces for the purposes of law enforcement," subject to narrow exceptions.

These three categories carry distinct implications for the allocation of responsibility between providers and deployers. In the first category, both the provider and the deployer violate the prohibition, because the mere capability of the system to exhibit the prohibited behaviour suffices to trigger the norm – regardless of who placed it on the market and for what purpose. In the second category, the deployer breaches the prohibition whenever the system is used for the incriminated purpose; the provider, by contrast, only violates the prohibition if it narrowly designed the system with this specific purpose in mind. In the third category, only the deployer can violate the prohibition, since the norm exclusively targets use.

For AI agents, these distinctions matter in a particular way. A general-purpose agentic system – released as a GPAI system without a designated prohibited purpose – can nonetheless fall within the scope of the prohibitions. In the first category, the provider faces liability simply by virtue of the agent's capabilities. In the second and third categories, the responsibility shifts to include concomitantly or exclusively the deployer who directs the agent toward the prohibited activity. Importantly, in all cases, it is enough for an AI agent, understood as an AI or GPAI system, to be released and to be used for the incriminated activity to trigger the prohibitions. Overall, concerning the provisions of the AI Act, providers or deployers cannot "hide" behind the non-specific, general-purpose capabilities of the agent.

This is partially different for the high-risk rules. The Act imposes transparency and accountability obligations for high-risk systems, but these are challenging to meet given the autonomy and complexity of agentic AI – but also their general-purpose character. Under the logic of the AI Act, inherited from product safety law, the system only qualifies as high-risk if the provider endowed it with the purpose intended to be used in this context. For example, Article 6(1) stipulates that certain AI systems qualify as high-risk if the "AI system is intended to be used as a safety component of a product, or the AI system is itself a product, covered by" the so-called "New Legislative Framework" for product safety regulation in the EU (Annex I Section A AI Act). These specific product safety rules typically require the product to be intended for the regulated purpose. Similarly, under Article 6(2) AI Act in conjunction with Annex III, systems in areas such as employment, certain parts of education, critical infrastructure or biometrics qualify as high-risk if they are "intended to be used" in these areas.

As discussed above (Section 3.2.2), this specific intention is different from the mere capability of being used for these activities. Hence, a GPAI system without a specific purpose designation does not qualify as high-risk. If providers want to exercise extra caution, they can exclude high-risk areas in the instructions for use.

The upshot of this categorical systematic distinction between specific high-risk purposes and general purposes is that generic AI agents, which are not endowed with a specific high-risk purpose, do not fall under the high-risk rules of the AI Act. However, if the deployer does use them in a high-risk context, Article 25(1)(c) AI Act kicks in. At least generally, the deployer, by changing the purpose from general to a specific high-risk setting, is transformed into a provider of what now becomes a high-risk system and this specific scenario. Hence, if an agent is used in an employment context, the

deployer – but not the provider of the original agentic system – becomes the provider of a high-risk AI system. We will discuss some nuances concerning this change of provider status below in the section on value chain (Section 3.4); before that, we turn to the content of high-risk rules and their meaning for AI agents.

The application of high-risk requirements under the AI Act to agentic AI systems also raises several substantive challenges. These relate, in particular, to provider obligations, record-keeping, transparency, human oversight, and robustness.

First, the provider obligations established for high-risk AI systems may prove difficult to apply where an agent is deployed for purposes unforeseen by the original provider. For instance, Article 10 requires that training, validation, and testing data sets be representative, relevant, and sufficiently complete. Yet if an agent autonomously pursues a new purpose not anticipated at the time of development, the original data set may no longer satisfy these criteria. This problem is compounded in complex value chains, where the entity that places the system on the market may differ from the entity that originally developed the underlying model – a challenge addressed in greater detail below (3.4).

Second, Article 12 on record-keeping requires high-risk AI systems to log their actions to facilitate accountability and traceability. For autonomous agents that make decisions independently, however, it may be technically difficult and conceptually opaque to record every step of the decision-making process. If every step is logged, this would also consume an enormous amount of storage and energy, making this doubtful from a sustainability perspective.

Third, Article 13 on transparency mandates that high-risk systems providers offer clear and comprehensible information about their functionality and decision logic to users and regulators. Given the dynamic and context-dependent nature of agentic AI, the achievement of meaningful transparency will require the development of robust explanatory frameworks.

Fourth, Article 14 on human oversight is of particular relevance. Since agentic systems operate autonomously, the maintenance of effective oversight becomes increasingly difficult as complexity scales. The very autonomy that defines agents thus stands in tension with the Act's oversight requirements.

Fifth, Article 15 emphasizes technical robustness and safety. For agentic systems, reliability is paramount: they must be capable of handling rare and unpredictable edge cases without system failure.

Overall, these provisions highlight both the ambition and the difficulty of the existing AI Act framework when applied to agentic systems. Importantly, a clear regulatory mandate already exists that specifies requirements applicable to AI agents as much as to any other AI system. In essence, the risks posed by AI agents do not generally differ from those posed by other AI systems in any substantial way. What does differ are the ramifications of agent autonomy, which means that higher thresholds will be needed to demonstrate the trustworthiness of these systems.

Beyond provider obligations, the AI Act also imposes significant duties on deployers of high-risk AI systems. Article 26(5), in particular, requires deployers to monitor the operation of such systems. For agentic AI, this obligation presents distinct difficulties: where an agent reasons and makes decisions at high speed and in parallel across multiple tasks, real-time human monitoring may become practically infeasible (Nannini et al., 2026). As we explore below, this challenge may necessitate the development of semi-automated monitoring frameworks that can keep pace with the operational tempo of autonomous agents.

### *3.4 Value chain considerations*

The AI Act allocates responsibilities along the AI value chain. As a baseline, the provider of a GPAI model bears the obligations set out in Articles 51–56. When a downstream entity modifies the model or system, or changes its purpose, however, the question arises whether that entity becomes a new provider – either of a GPAI model or of a high-risk AI system. This question is of particular importance for agentic AI, where an entity may wrap a foundation model in an agentic scaffolding that substantially alters the system's capabilities and risk profile.

**Becoming a new provider under Article 25**

Article 25(1) of the AI Act stipulates that any distributor, importer, deployer, or other third party shall be considered a provider of a high-risk AI system in certain circumstances. Two pathways are especially relevant for agentic AI.

First, under Article 25(1)(b), an entity that makes a substantial modification to a high-risk AI system already placed on the market becomes a new provider. By analogy, as the European Commission's guidelines on GPAI models suggest (European Commission, 2025b; see also Hacker & Holweg, 2025; Ebers & Penagos, 2026), an entity that modifies a GPAI model through agentic scaffolding may become a new GPAI model provider if – and this is a big if – the scaffolding alters the model itself (and not only the system) in a substantial manner.

Second, and more consequentially, Article 25(1)(c) AI Act provides that an entity becomes a high-risk system provider if it "modif[ies] the intended purpose of an AI system, including a general-purpose AI system, which has not been classified as high-risk and has already been placed on the market or put into service in such a way that the AI system concerned becomes a high-risk AI system in accordance with Article 6." This provision is central when an agent built on a GPAI model is deployed in a high-risk domain such as medical diagnostics, employment, or critical infrastructure management.

**The doctrinal problem under Article 6(1)**

However, a doctrinal problem arises in this context. Article 25(1)(c) refers to Article 6, which sets out the conditions under which an AI system qualifies as high-risk. Article 6(1) refers to the New Legislative Framework (NLF) of product safety law – sectoral regulation such as the Medical Devices Regulation (MDR) or the Machinery Regulation (Annex I); Article 6(2) relates to activities enumerated in the AI Act itself (Annex III), such as employment or critical infrastructure. Under Article 6(1) specifically, a system is high-risk where two cumulative conditions are met: (a) the system is intended to serve as a safety component of a product, or is itself a product, covered by the Union harmonization legislation listed in Annex I; and (b) the product is required to undergo a third-party conformity assessment under that legislation.

Yet, the interplay with Article 25(1)(c) creates a significant doctrinal tension with established NLF jurisprudence. Not all NLF norms recognize a change of purpose by a downstream actor as an activity changing that entity's status to manufacturer. Whether a product falls under NLF legislation – such as the MDR or the Machinery Regulation – and must undergo third-party conformity assessment has typically depended on the intended purpose as defined by the *original* manufacturer, not the actual use by downstream actors. The CJEU confirmed this principle for the predecessor Medical Devices Directive,[3] and the BGH has likewise held that the classification of a dual-use product as a medical device depends on the subjective determination of the manufacturer, provided that the stated purpose is unambiguous, non-arbitrary, and a non-medical use is plausible[4] (cf. Ebers & Streitbörger, 2024; Noller & Rappenglück, 2026). Thus, under this established case law, even if an agent is in fact used for medical diagnostics, it would not necessarily become subject to the MDR – and hence would not trigger high-risk classification under Article 6(1) – as long as the original model provider did not designate the system for a medical purpose.

A similar logic governs Regulation (EU) 2023/1230 on machinery. Article 3 defines „machinery" and related products by reference to their intended function, and the manufacturer must determine the intended use and reasonably foreseeable misuse as part of the conformity assessment and risk analysis (Art. 6(4), 8, Annex III B(1)).

More recent regulatory developments align NLF doctrine more closely with Art. 25(1)(c), though. Noller and Rappenglück rightly note that the MDR predecessor directive did not incorporate the concept of reasonably foreseeable misuse, and that the explanatory Recital 6 in Amending Directive 2007/47, on which the CJEU relied, clarified that general-purpose software qualifies as a medical device only if the manufacturer so intends. This has changed with the MDR. More specifically, Article 16(1)(b) MDR now contains a rule that mirrors Art. 25(1)(c). If another person changes the purpose of a device already placed on the market, that person assumes the responsibilities of the manufacturer. This seems to capture cases in which the deployer pivots a general-purpose AI system toward medical use if the original agent qualifies as a "device." While the GPAI system is not a medical device, Art. 16(1)(b) does not presuppose this, demanding only a device plain and simple. Software clearly qualifies as a device (cf. Art. 1(1) MDR). Hence, Article 25(1)(c) extends this novel NLF doctrine to the AI Act. The exact same change-of-purpose rule is found in the in the *in vitro* diagnostics regulation (IVDR), too (Art. 16(1)(b) Regulation (EU) 2017/746).

A direct equivalent to Art. 16(1)(b) MDR and Art. 16(1)(b) IVDR is, however, missing in other NLF regimes, such as the Machinery Regulation (MR). Art. 18 MR covers cases of substantial modification: The person making a substantial modification to machinery or a related product assumes the obligations of the original manufacturer. Article 3(16) MR, in turn, defines a substantial modification as a change not foreseen by the original manufacturer which adds or increases the risk and necessitates the addition of protective devices or measures for mitigation. One might extensively read Article 18 MR to also cover cases of substantial modification to *safety components*, and to also include cases in which the *purpose* of a general-purpose software was changed. This is supported by Article R6 of Annex I of Decision No 768/2008/EC, a decision concerning products regulated under the NLF. Said article stipulates that importers or distributors assume the responsibilities of manufacturers if the modify a product such that the compliance with the applicable requirements may be affected.

However, there is a significant degree of uncertainty concerning such an extensive reading of Art. 18 MR. First, other NLF regimes, such as the MDR and the IVDR, distinguish between substantial modification and change of purpose (Art. 16(1)(b) vs (c) MDR/IVDR), just like the AI Act in Article 25(1).

---

[3] CJEU, Case C-219/11, Brain Products, para. 16-17; Case C-329/16 (Snitem), para. 24.
[4] BGH, Case I ZR 53/09 , Messgerät II, NJW-RR 2014, 46.

Second, the mentioned provision of Decision No 768/2008/EC only covers modifications by importers and distributors, not by users or deployers. Third, the CJEU case law on the relevance of the original manufacturer's intent, even if handed down in the context of the MDR, suggests that original intent remains dispositive in the absence of a clear legal provision to the contrary. The Blue Guide, a guideline for the application of all NLF rules, now contains general provisions concerning modifications (incl. change of purpose) by downstream actors, and the ensuing assumption of manufacturer obligations by them (European Commission, 2022, Section 2.1, p. 15, and Section 3.1, p. 38; see also Nannini et al., 2026). However, the Blue Guide constitutes mere soft law and cannot, on its own, overrule CJEU case law or make up for a lack of specific normative provisions in the acts themselves.

The exact same issue arises with respect to GPAI agents repurposed as safety components for toys (Art. 8 Directive 2009/48/EC) or recreational craft and personal watercraft (Art. 11 Directive 2013/53/EU), which essentially mirror Art. 18 MR.

Hence, the relevance of the original manufacturers intent remains a doctrinal problem in some areas of the NLF in which change of purpose by users or deployers is not explicitly tied to a change to manufacturer status; concerning medical devices and IVMD more specifically, however, Article 16(1)(b) MDR and Article 16(1)(b) IVDR arguably render the CJEU case law, and the implementing decision of the BGH, moot and solve the doctrinal friction for these areas.

**Article 25 as lex specialis**

If one were to apply the established NLF doctrine about the relevance of the original manufacturer's intent without modification, Article 25(1)(c) would be deprived of any meaningful scope of application with respect to significant parts of legislation covered under Article 6(1). A deployer who repurposes a GPAI model for use in a machinery, toy, or other NLF context still subject to that doctrine would never trigger high-risk obligations, because the original manufacturer's intended purpose – not the deployer's – would remain decisive for the NLF classification. This result would produce substantial protection gaps precisely in those domains – such as industrial and toy safety – where high-risk regulation matters significantly.

In our view, Article 25 must therefore be understood as lex specialis relative to the specific doctrinal contours of sector-specific product safety regulation. The provision was designed to capture precisely those situations in which downstream actors redirect a system toward high-risk uses that the original provider did not foresee or intend. If the deployer, for example the agent builder or agent user, changes the intended purpose from a general-purpose application to a high-risk use case, Article 25(1)(c) demands that the system (e.g., agent) be treated as high-risk. More specifically, the AI system becomes high risk provided that all substantive conditions for the high-risk activity – whether under Annex III or the NLF legislation in Annex I(A) – are met, with the exception of the requirement that the original manufacturer specified the high-risk purpose (if applicable). As a result of the lex specialis character of Art. 25(1)(c) AI Act, it suffices that the deployer effectuates this change of purpose – even if, under the specific NLF legislation (e.g., the MR or the Toy Directive), such a change by a non-manufacturer would ordinarily not bring the system within that legislation's scope (cf. also Noller & Rappenglück, 2026, para. 17).

The consequence is notable: an agent deployed in safety-critical industrial machinery settings constitutes a high-risk AI system under Article 25(1)(c), even if it may not qualify as a regulated device under the MR or similar NLF acts, unless the CJEU changes its case law. The AI Act and the NLF framework may thus, in some cases, part ways – even if this divergence was not originally intended by the legislature.

**The Distinction Between Article 6(1) and Article 6(2)**

It bears emphasis that this doctrinal problem arises only under Article 6(1), which ties high-risk classification to NLF legislation and its manufacturer-centric logic. Under Article 6(2), by contrast, high-risk classification follows from Annex III, which lists specific use cases – such as biometric identification, critical infrastructure management, or employment-related decisions – that are not governed by NLF legislation at all. Annex III sectors appear in that annex, and not in Annex I, precisely because no harmonized NLF legislation exists for them. Although Annex III also relies on the concept of intended purpose, Article 25(1)(c) makes explicit that, in value chain constellations, the intended purpose of the deployer is the relevant benchmark. Accordingly, a deployer may become a high-risk provider by redirecting a GPAI-based agent toward an Annex III use case, without the doctrinal friction that arises under Article 6(1).

**Regulatory Implications**

The divergence between the AI Act and NLF-based product safety regulation creates a distinctive compliance dilemma (see again also Ebers & Streitbörger, 2024; Noller & Rappenglück, 2026). An agent builder who uses a general-purpose chatbot for NLF purposes would, under the AI Act, become a provider subject to the full suite of high-risk obligations, while under the relevant NLF regime, the system may not be classified as a regulated device because the original manufacturer (e.g., OpenAI) did not endow it with a relevant purpose. This produces a situation in which the deployer faces potential liability under one regulatory regime but not under the other.

A key problem arises here: the new high-risk provider needs to comply with Chapter III obligations, but does not have access to many of the information pieces needed for compliance, because these reside with the GPAI model provider. Article 25(2) seeks a solution for this by stipulating collaboration duties and access rights. However, in the trilogue negotiations, GPAI model providers were removed as addressees of this right; it is now limited to high-risk GPAI system providers. In effect, this means that new providers are covered, but it is almost impossible for them to comply with the obligations they are now faced with, such as data governance under Art. 10 AI Act, when the actual data processing and model training was conducted by the GPAI model provider (Hacker, Kilian & Costas, 2025).

### 3.5 New Regulation of Agents as "GPAI Agents"?

A question that has received insufficient attention in the literature so far is whether some very basic AI agents could – or should – be treated as general-purpose AI (GPAI) models under the AI Act, which would trigger the obligations set out in Articles 53 and 55 AI Act, or even necessitate a different category. The Act imposes transparency, documentation, copyright-compliance, and downstream-information duties on providers of GPAI models (Article 53) and adds further requirements – model evaluation, systemic-risk assessment, incident reporting, and cybersecurity obligations – for providers of GPAI models with systemic risk (Article 55). These obligations apply since August 2, 2025, and the Commission's July 2025 Guidelines on the Scope of GPAI Model Obligations clarify that a model qualifies as "general-purpose" where it was trained with more than $10^{23}$ FLOPs and is capable of generating language, text-to-image, or text-to-video outputs (European Commission, 2025b, para. 17). The systemic-risk presumption is presumptively triggered at $10^{25}$ FLOPs.

At present, this framework is directed at foundation models, not at agents built on top of them. The July 2025 Guidelines from the European Commission and current literature (Blum & Rappenglück, 2024; Hacker & Holweg, 2025; see also Schwartmann & Zenner, 2025) suggest that techniques such as retrieval-augmented generation, prompt engineering, orchestration, and tool-calling do not constitute significant modifications to the underlying model and therefore do not make the entity that employs these techniques a GPAI model provider (e.g., under an analogy to Art. 25(1)(b) AI Act). Only modifications that alter the model weights above a compute threshold of roughly one-third of the original training compute, or a consequence scanning that reveals significant risk shifts (Hacker & Holweg, 2025), can trigger provider status. This means that most agent builders will not fall within the scope of Chapter V of the AI Act at all; the GPAI obligations remain with the upstream model provider.

Yet, this settled picture may, in the longer run, be too simple. One could conceive of a category of "base" or "foundation" agents: agentic systems that are themselves sufficiently general-purpose and widely deployable to warrant treatment analogous to GPAI or "foundation" models, even if they do not cross the compute and modification thresholds currently discussed. In the same way that a GPAI model serves as a foundational component that can be integrated into many downstream AI systems, such a base agent could serve as a foundational agentic framework that is integrated into a variety of downstream applications and use cases. Three examples illustrate this possibility.

First, consider a "person evaluator" agent. This would be an agentic system built with novel scaffolding, memory structures, and tool-use capabilities that is designed to assess individuals across a range of contexts, and directly execute a differentiation solution. Such a system might be deployed in employment screening, educational admissions, and medical triage alike. Each of these downstream use cases would independently qualify as high-risk under Annex III of the AI Act. The agent itself, however, is designed as a general-purpose evaluation module that can be plugged into different domains.

Second, imagine a "coding agent" that combines a foundation model with specialized scaffolding for code generation and execution. If such an agent is designed to code, test, and deploy IT security updates autonomously on laptop computers and tablets worldwide, its scope of impact is vast. Critically, the agent's capacity for autonomous real-world action – the hallmark of agentic AI – amplifies the risk profile beyond what the underlying model alone would present.

Third, recent practice and academic literature suggests that LLMs are increasingly used as an orchestration layer. These systems are agents coordinating a whole network of subagents, routing tasks to these more specialized agents, supervising their work, and collecting and presenting results (Ding et al., 2024; Shnitzer et al., 2023; Webb et al., 2025; Yang et al., 2025). These larger orchestration-level agents act as coordinators and as an interface to the human user, while smaller models powering the specialized subagents (Belcak et al., 2025; Ngo-Ho et al., 2026). This avoids excessive token usage by large models (Bergemann et al., 2025; Erdil, 2025) and may lead to what has recently been termed "domain-specific superintelligence" encapsulated by the specialized agents (Belova et al., 2026). The orchestration agent, hence, supervises a whole fleet of subagents; it has, therefore, significantly more leverage to propagate risk into a system or agentic network than a single agent.

Currently, such agents qualify as GPAI systems (see Section 3.2.2), and are hence generally exempt from specific systemic risk identification and mitigation obligations (Art. 55) even though the agents may be used by different downstream actors in various contexts, including such with high stakes. Against this background, should the law be updated to include a specific risk identification and mitigation regime for base agents, for example under a new Chapter for "GPAI agents"?

**The Case for Regulating "Base Agents"**

Several arguments support the introduction of a GPAI-like regime for base or foundation agents, i.e., a GPAI agent framework.

The most important argument is one of regulatory and economic efficiency. If a base agent is widely deployed downstream, it is more efficient to address the relevant risks once, at the source, rather than to require each downstream deployer to conduct independent risk assessments and compliance measures for the agent-level capabilities. This mirrors the logic of the GPAI model framework itself: the layered structure of the AI Act rests on the premise that shared foundational components should be regulated at the level at which they are provided, so that the benefits of compliance cascade down the value chain (Hacker et al., 2023; Wachter, 2023).

Furthermore, the scaffolding layer of an agent may itself introduce new risks that are not captured by the regulation of the underlying model. Agent scaffolding determines how the model interacts with tools, accesses memory, plans multi-step actions, and manages the boundaries of its own autonomy. These architectural choices can materially affect the risk profile of the resulting system. If the underlying model is subject to Articles 53 and 55, but the scaffolding that transforms it into an autonomous agent is not, a regulatory gap may arise (Oueslati & Staes-Polet, 2025).

A third consideration is that the underlying model may itself not qualify as a GPAI model, or at least not as a GPAI model with systemic risk. Agents can be built on top of smaller, specialized models that fall below the Commission's $10^{23}$ FLOPs threshold and therefore do not trigger any Chapter V obligations; or below the 1025 FLOPs threshold for systemic risk. In such cases, the GPAI model regime does not apply at all, or at least Art. 55 AI Act does not apply, even though the resulting agent – through its scaffolding, tool access, and deployment breadth – may present risks comparable to those of a system built on a GPAI model with systemic risk. This gap would only widen as efficient, sub-threshold models become more capable. However, some of these gaps might be plugged by an analogy to Art. 25(1) AI Act (see below, 3.5.2).

Fourth, the downstream use cases of a base agent may not necessarily qualify as high-risk under Article 6 and Annexes I and III of the AI Act. While a person evaluator typically falls under Annex III and is subject to Art. 9 AI Act's risk management provisions, a coding agent that deploys security updates on personal computers, for instance, does not map onto any of the Annex I or Annex III categories. The high-risk classification regime thus does not apply, and the GPAI model obligations, if any, rest with the upstream model provider rather than the entity that assembled the agent. The result is that neither the system-level nor the model-level provisions of the AI Act adequately capture the agent-specific risks in coding contexts.

If a GPAI agent framework were to be introduced, it would require a provision analogous to Article 25 of the AI Act – the article that allocates responsibilities along the AI value chain for high-risk systems. Just as Article 25 provides that a third party who substantially modifies a high-risk AI system or changes its intended purpose becomes the provider of that system, an "Article 25 for agents" would stipulate that the entity who assembles a GPAI model into a base agentic system with general-purpose capabilities becomes the provider of the resulting "GPAI agent." That entity would then bear obligations comparable to those in Articles 53 and 55, adapted to the specific risk profile of agentic systems, for instance, with respect to tool-use safety, autonomy boundaries, and the cascading consequences of multi-step action execution.

**The Case Against a New GPAI Agent Framework**

At the same time, there are substantial reasons for caution. The weightiest counterargument is that the core of the risk typically originates in the underlying GPAI model, and this risk is already covered by the GPAI model provisions. The agent, in this view, is merely an extension of the model within a broader system, and the appropriate regulatory layer is therefore the system-level regulation of the AI Act – in particular, the high-risk classification regime of Chapter III. The agent, typically, does not create fundamentally new capabilities that are independent of the model; it channels the model's capabilities in specific ways. Even orchestration-level agents are ultimately systems depending on their underlying models for performance; they do influence numerous other agents, but this is again a specific risk that model providers have to foresee and mitigate under Article 55. Agent-level regulation would thus be duplicative. Furthermore, to the extent that the agent does introduce significant novel risks vis-à-vis the underlying model, the agent builder may actually become a new model provider (Article 25(1) AI Act), and be subject to Chapter V AI Act, even if the one-third FLOP threshold is not reached (Hacker & Holweg, 2025). This may even mean that, if only the modification makes the model cross the systemic risk threshold, the modified version qualifies for Art. 55 AI Act (cf. European Commission, 2025b). However, this remains unclear absent a final judgment by the CJEU, particularly since it may be argued that the novel risks do not enter at the model but only at the system level (Section 3.1), and can, therefore, not be addressed by Chapter V under the core regulatory architecture of the AI Act.

A related, second concern is regulatory complexity and cost. The AI Act already imposes a multi-layered compliance architecture on providers and deployers, as well as importers and distributors, of models and systems. A new layer of "GPAI agent" obligations would add significant compliance burden, especially for the many small and medium-sized enterprises that build agentic systems on top of commercially available foundation models. As the Commission's 2025 Digital Package on Simplification signals, there is an active policy concern about the cumulative weight of AI Act obligations. Adding a new regulatory category for agents, hence, does not seem likely to be politically feasible at this moment – even though this pragmatic consideration does not, of course, substantively speak against such novel obligations.

However, third, important aspects of agent-specific risk are already addressed by adjacent regulatory instruments. The Cyber Resilience Act (CRA), which entered into force on December 10, 2024, and whose main obligations will apply from December 11, 2027, imposes mandatory cybersecurity requirements on manufacturers of products with digital elements – a category that can encompass software agents that are placed on the market as products (Novelli et al., 2024a). The NIS 2 Directive, transposed in several Member States by late 2025, strengthens cybersecurity governance and incident-reporting for essential and important entities. Together, the CRA and NIS 2 cover a significant portion of the cybersecurity and operational resilience risks that agents pose, particularly in critical infrastructure and IT security contexts. In addition, civil liability rules – both under general tort law and under the revised Product Liability Directive – provide ex post corrective mechanisms that can address harms caused by agentic systems without the need for a dedicated ex ante regulatory framework within the AI Act (cf. de Bruyne et al., 2023; Hacker, 2023; Novelli et al., 2024a; Wagner, 2023).

**Assessment**

On balance, in our opinion, there is no major urgency for regulatory reform at the present moment. The existing framework of the AI Act – the GPAI model obligations in Articles 53 and 55, the high-risk system requirements of Chapter III, and the value-chain responsibilities of Article 25 – combined with the CRA, NIS 2, and civil liability, provides a workable regulatory architecture for addressing agent-related risks. The concept of a "base agent" that is both sufficiently general-purpose and sufficiently independent of its underlying model to warrant a separate GPAI-like classification remains, at this stage, more theoretical than empirically grounded.

That said, the landscape is evolving fast. As agents become more autonomous, more widely deployed, and more architecturally independent of their underlying models – for instance, through multi-model orchestration, persistent memory, and direct real-world tool access – the case for a dedicated "GPAI agent" category may become stronger. The AI Act's built-in review mechanism provides a natural occasion for reassessment. At a minimum, the Commission's future guidelines and the ongoing development of harmonized standards should explicitly address agent-specific risks and clarify how the existing provisions apply to agentic systems. This topic deserves close monitoring and should feature prominently in the AI Act's scheduled review process.

### 3.6. *Policy implications: Regulating AI Agents in the Context of the AI Act*

What transpires from the preceding discussion is that a key challenge in regulating agentic AI arises from the diversity of contexts in which such systems are deployed. Agents perform vastly different functions across domains, each with its own scope and level of risk. Understanding the role of context is therefore essential to crafting effective governance frameworks.

For example, when AI tools are used internally for low-risk purposes such as document retrieval, they generally pose minimal risks under the EU AI Act and have no specific regulatory mandate in the AI Act beyond AI literacy (Art. 4). By contrast, AI agents deployed in coding or customer service chatbots raise questions of transparency and accountability (Art. 50 AI Act), though their risks to fundamental rights remain limited, except in specialised domains such as health care where sensitive data are involved. When agentic AI is applied to decision-making workflows in areas such as human resources or financial services, the risks mirror those of non-agentic AI; here, the risk stems from the context of use rather than from the underlying model architecture (Annex III). However, in high-stakes applications involving physical systems–such as autonomous driving, cobots, drones, and autonomous delivery vehicles–agents must develop situational awareness to ensure safety in environments shared with humans. These risks relate primarily to physical safety and are addressed under existing regulatory frameworks, such as Annex I of the AI Act and the sectoral acts invoked there. While many of these sectoral regulations await an "AI update," analysis of and suggestions for these specific sectoral regimes transcend the scope of this paper.

| Use Case | Example Applications | Risk Level under EU AI Act | Primary Risks/Concerns | Regulatory Implications |
|---|---|---|---|---|
| **Internal, low-risk use** | Document retrieval, internal assistance | Minimal risk (if no direct interaction with humans) | None that are significant | No specific regulatory mandate required |
| **Customer-facing conversational AI** | Coding or customer service chatbots | Limited risk (context-dependent) | Transparency and accountability concerns; potential issues in specialised domains (e.g., health care with sensitive data) | May require transparency safeguards; limited obligations except where sensitive data is used |
| **Decision-making workflows** | Human resources, certain financial services | High risk (context-dependent, Annex III) | Bias, fairness, explainability–risks mirror those of non-agentic AI | Risk assessment depends on use context rather than model architecture |
| **High-stakes physical applications** | Autonomous driving, cobots, drones, autonomous delivery vehicles | High risk (context-dependent, see Annex I) | Physical safety, situational awareness, human interaction risks | Covered under existing safety regulations (e.g., Annex I of EU AI Act and sectoral acts) |

**Table 2: Agentic AI use cases, risks, and regulatory coverage**

The results of this analysis indicate that evaluating the abstract performance of individual model components is insufficient to fully understand overall system behaviour in the field. Compliance, therefore, cannot be assessed in a purely summative manner. Contextual evaluation is essential: agents must be assessed within the environments in which they are implemented. The implication is clear–agents should be treated as complex, heterogeneous systems that must be tested within their operational contexts. Their risks depend on the use case and deployment environment, not primarily model architecture. Consequently, the GPAI-model-oriented approach adopted by the AI Act for frontier LLMs is not directly applicable to agentic AI.

Building on this and the detailed legal evaluation in Section 3, we make three key proposals to enable an effective contextual regulatory regime under the AI Act. They pertain to the content of high-risk rules, more specifically monitoring duties (Proposal 1); the classification as a high-risk provider under the value chain rules (Proposal 2); and the collaboration with the provider of the underlying GPAI model (Proposal 3).

**Proposal #1** [see Section 3.3.3.2]: For autonomous agents in high-risk contexts, the frequency and scope of human oversight should be subject to mandatory specification, accompanied by a corresponding documentation mandate. The current provisions – Article 26(5) and (6) – address deployer oversight in general terms but do not account for the particular demands that agentic autonomy places on monitoring regimes. We therefore recommend the addition of specific rules or regulatory guidance under these provisions that establish a mandatory oversight protocol and schedule tailored to autonomous agents. Such a protocol should define minimum review intervals, specify the scope of oversight activities, and require structured documentation of oversight outcomes.

**Proposal #2** [see Section 3.4.3]: Guidelines or, preferably, an amendment to the AI Act should specify that Art. 25(1)(c) AI Act applies even if the intended purpose of the original system or model provider is the sole criterion under legislation referenced in Annex I A. This covers cases in which the deployer decides to use an AI system, such as an agent, in a high-risk area defined by Annex I A, such as machinery or toy settings. In this case, the deployer changes the originally intended purpose of a non-high-risk system (e.g., an agentic GPAI system) to the high-risk use case. Even though the intention of the original provider is arguably dispositive under the sectoral acts referenced by the AI Act in Annex I A (with some exceptions, e.g. the MDR and the IVDR), the change of purpose by the deployer must trigger the change of status – from deployer to provider – under Article 25(1)(c) AI Act. Only in this way, the application of the specific high-risk rule of the AI Act, with contextual risk management, performance, cybersecurity and robustness obligations, can be triggered.

**Proposal #3** [see Section 3.4.5]: The collaboration duties under Art. 25(2) should be extended to GPAI model providers. The mere duty to conclude a written agreement under Art. 25(4) AI Act does not suffice as it does not contain a substantive collaboration duty (Hacker, Kilian & Costas, 2025; Gössl, 2024, para. 56). Only an extension of Art. 25(2) ensures that downstream deployers who become providers (because they use agents in high-risk contexts) then have a right to obtain access to the information necessary for compliance with the high-risk rules they are now subject to (e.g., concerning data governance for the underlying model, Article 10 AI Act).

## 4 Agentic autonomy and contract law: Facilitation and guardrails

The AI Act is, by far, not the only regulatory layer extending to agents. Just like tort law, the AI Act deals with agents, and AI models and systems more broadly, from the point of view of *risk regulation*, specifically concerning negative externalities arising from agents. The specific autonomy of and automated task execution by agents, however, intimately relates to another legal field: contract law. If agents send emails, order goods, and exchange content, the question arises whether such agentic output rises to the level of legally valid and binding contractual declarations.

This is not a new question, and we cannot, for reasons of scope, provide a full overview of contractual problems in agent settings (see also, e.g., Busch, 2025; Kolt, 2025; Wendehorst, 2024). However, we shall discuss some key points of contention that arise from the autonomy contemporary and future agents exhibit and that have not been sufficiently captured by the literature, so far. Indeed, contract law has grappled with automated and autonomous means of negotiating, concluding, executing, breaching, and enforcing contracts for a long time (see only Bellia, 2001; Hacker, 2020, 610 et seqq.; Hacker & Grundmann, 2017; Kidd & Daughtrey, 1999; Lerourge, 2017; Oliver, 2021; Scholz, 2017; Savelyev, 2017; Sklaroff, 2017; Weitzenboeck, 2001; United Nations, 2005). These questions have risen to prominence again with the advent of contemporary generative AI and agents (European Law Institute, 2025; Migliorini, 2024; Poncibò, 2023; United Nations Commission on International Trade Law, 2024; Wendehorst, 2024). Contract law rules, broadly speaking, deal primarily with certain risks for counterparties (as opposed to those for third parties) arising from the collaboration structured by the contract, and the exchange of information and declarations via agents (e.g., errors, breaches of internal agency constraints).

Increasingly, disputes around chatbots and other AI-based assistants show that automated outputs can indeed trigger legally relevant reliance and, in some settings, contract-law consequences. In Moffatt v. Air Canada, the Civil Resolution Tribunal in British Columbia held Air Canada liable for negligent misrepresentation after its website chatbot gave misleading information about bereavement fares; the Tribunal treated the chatbot as part of the airline's website and rejected Air Canada's attempt to cast the chatbot as a separate "agent" that would break attribution (St-Hilaire, 2025). A separate, widely reported incident involved a Chevrolet dealership website chatbot that assented to an obviously absurd "offer" after user prompt manipulation. This illustrates how interface-level attacks can induce exchanges, legally binding or not, in consumer-facing dialogues which are not in the interest of the AI deployer (Notopoulos, 2023).

These episodes do not determine EU outcomes, yet they clarify a general point that legal theory has addressed for decades: electronic agents can participate in offer and acceptance, and private law can attribute their statements to the party that deploys them (Weitzenboeck, 2001; Hacker & Ebert, 2025; Wendehorst, 2024).

### *4.1 Baseline in EU Law: Electronic Contract Formation*

EU law contains a narrow, functionally oriented harmonization for electronic commerce. Articles 9–11 of the E-Commerce Directive (Directive 2000/31/EC) require Member States to ensure that contracts may be concluded by electronic means and to regulate certain information duties and technical steps that structure electronic ordering processes (Rösler, 2012). The Directive supports electronic contract formation, yet it does not supply a general attribution regime for fully autonomous systems that substitute for a natural person at the moment of contract conclusion.

Doctrinally, Member State systems already rely on attribution techniques and objective formation models that allow contract law to function despite imperfect psychological assumptions. Representation doctrine often proceeds via legal fictions around assent and attribution, with substantial debate on the doctrinal details (Friedman, 2003; Hacker, 2020; Kötz, 2017; Sage, 2023 Wendehorst, 2024). That insight aligns with earlier scholarship on "electronic agents," which argues that contract formation can proceed without a natural person who reviews each discrete communicative act, provided attribution rules allocate responsibility to the operator (Weitzenboeck, 2001). As a consequence, the European Law Institute's Model Rules on Digital Assistants for Consumer Contracts rightly rest on a risk-allocation premise: acts of a digital assistant are attributed to the person who uses it, subject to explicit limitations, and consumer protection must remain effective even when automation displaces direct human participation (European Law Institute, 2025). The Model Rules combine private-law attribution rules with design requirements and emphasize that automation can generate legal consequences for humans even when they did not participate in the specific act (European Law Institute, 2025).

*4.2   Persistent Blind Spots: Error and Ultra Vires Acts*

Two gaps remain salient, however, once AI agents act autonomously, at scale, and without contemporaneous supervision.

First, error rules and remedies vary across Member States. Member State laws may differ on the consequences of an agent's mistake, such as an incorrect order, an inaccurate price statement, or an unintended acceptance. These divergences matter more when AI systems produce outputs through probabilistic processes that principals cannot fully anticipate. International harmonization texts already address adjacent issues. The United Nations Electronic Communications Convention recognizes contract formation through automated message systems even when no natural person reviewed each individual action, and it contains a dedicated rule on input error in communications with such systems (United Nations, 2005). These provisions treat automation-compatible error allocation as a central design issue, not as a rare anomaly. Ultimately, errors will be attributable to the principal.

Ultra vires transactions and the absence of clear ex ante limits pose a second problem. Agency law often lacks bright-line, general prohibitions on categories of acts by agents. Sometimes, however, such dedicated limits do exist. For example, under German business law, a human general agent does not possess authority to conduct a sale of the principal's entire enterprise, or to sell immobile property, such as real estate (§ 49 HGB). Internal limits on the mandate can exist (e.g., an explicit instruction to the agent not to engage in transactions above a certain financial threshold), yet third-party reliance and appearance-based doctrines can frustrate purely internal risk controls. Again, in business contexts, internal restraints on a director's power of agency are void concerning external contracting (see, e.g., § 37 GmbHG).

*4.3   Policy implications: An EU Regime for Agentic Systems in Contracts*

These discussions and gaps support the development of EU-wide rules tailored to agentic or other electronic systems, potentially as an optional regime that parties may choose. Such a regime could complement the E-Commerce Directive's electronic formation provisions (Directive 2000/31/EC, 2000) and align with broader international efforts toward automation-friendly contract law. UNCITRAL adopted the Model Law on Automated Contracting in July 2024 with the explicit aim of facilitating automation in contract formation and performance, with explicit reference to AI techniques, "smart contracts," and machine-to-machine transactions (UNCITRAL, 2024).

In this context we make three further proposals to facilitate agentic contracting. While the AI Act related proposals are designed to ensure effective oversight (regulation), this new set of policy suggestions aims to lay the ground for the effective use of agents' autonomy in market transactions (market-enabling law):

**Proposal #4:** General permission and non-discrimination for autonomous contract conclusion: The regime should clarify that contract validity and attribution do not fail solely because no natural person reviewed the specific agent action (see also Busch, 2025). This position matches the policy of the Electronic Communications Convention and the thrust of the UNCITRAL Model Law (UNCITRAL, 2024; United Nations, 2005). Attribution should default to the principal that deploys the system, subject to exceptions for counterparty bad faith and for evident excess of authority (European Law Institute, 2025; Hacker & Ebert, 2025).

**Proposal #5:** A statutory list of non-delegable or presumptively ultra vires transactions. Given the potential of manipulation of the agent, e.g., by prompt injection (Beurer-Kellner et al., 2025; Dziemian et al., 2026; Wang et al., 2026), the regime should define transaction types that no agent may conclude, or that require heightened safeguards. Such categories could cover dispositions of core assets, structural corporate acts, and commitments above standardized value thresholds (e.g., exceeding 1 Mio. €, depending on the sector). Clear ex ante lines reduce uncertainty for counterparties and protect principals against potentially catastrophic commitments that stem from system error or manipulation.

**Proposal #6:** An "authority traffic light" with machine-readable attestations. Beyond these enumerated and certain clearly communicated ultra-vires acts, we suggest a generally "externally unrestrictable authority." Under this model, internal limits on authority would not affect, i.e., not be binding on, third parties (see also Schirmer, 2016, 664; Grundmann & Hacker, 2017, 283); the binding effect of agentic declarations would fall away only in cases of recognized or evident excess of authority, i.e., if the counterparty knows or should have known about the limitations (Hacker, 2020, 611-12).

This approach can be analogized to common statutory constraints on external limits of representation in commercial and corporate law.[5] An EU regime could implement this concept through standardized grades of authority that principals select ex ante, paired with salient disclosure. A colour code ("traffic light regime") could communicate authority levels to humans, while standardized metadata or attribute certificates could communicate the same limits to automated counterparties (Sorge, 2006, 61 et seqq.). For example, the Air Canada chatbot could have communicated, within such a system, that it can be used for information and regular ticket purchase, but does not have authority to grant rebates or special conditions. This would have put the counterparty on notice, and any declarations concerning rebates would have been null and void.

Such standardization responds to a central weakness of free-form chatbot text: text alone cannot serve as a reliable representation of authority under adversarial pressure (Bundesamt für Sicherheit in der Informationstechnik, 2025, 24 et seqq.). Structured attestations would shift the legally relevant signal away from the conversational layer and toward machine-readable and machine-verifiable authorization statements. This, at a minimum, is needed for future machine-to-machine communications and contracts.

---

[5] In German law, for example, this would correspond to § 50 HGB, § 37(2) GmbHG, and § 82(1) AktG.

# 5 Conclusion

Our discourse has highlighted the level of autonomy as the defining difference between "traditional" AI systems and AI agents. From a legal perspective, there are at least two sides to the autonomy AI agents are increasingly endowed with. On the one hand, potential negative externalities, ranging from discrimination to cybersecurity issues and loss of control, need to be addressed via comprehensive risk regulation. This is the task of the AI Act, jointly with adjacent regimes such as the CRA, NIS 2, the GDPR, and product liability. On the other hand, the positive side of autonomy, such as automated task routing, solution, and execution, needs to be supported by dedicated contract law rules that confer validity on agentic declarations, but also impose minimum guardrails against errors and ultra vires acts.

We argue that the AI Act captures many, but not all, elements of the specific autonomy risks that agents pose. Its Article 3(1) highlights that an "AI system [...] is designed to operate with varying levels of autonomy". Indeed, agents that autonomously plan, reason, and execute tasks present perhaps the current frontier level of autonomy. Nonetheless, AI agents differ fundamentally from general-purpose AI models. Although they rely on GPAI models for reasoning and communication, they do not generally meet the GPAI definition in Article 3 of the EU AI Act because they are built for and operationalized in specific, context-dependent uses rather than broad, cross-domain functionality. Their inherently contextual operation calls for a regulatory approach that assesses risk based on deployment environment rather than underlying architecture. Agentic systems should therefore be regulated as domain-specific applications whose oversight obligations vary with context and autonomy, not as GPAI models – even though evolving agentic architectures, including orchestration agents, may eventually necessitate the creation of a novel category of "GPAI agents".

For the moment, however, the key feature that distinguishes agentic AI is a higher degree of autonomy. Because autonomous agents can act without direct human intervention and supervision, they introduce unique governance challenges. In high-risk settings–such as finance, employment, healthcare, or autonomous vehicles–mandatory oversight protocols are essential. These should clearly define monitoring scope, frequency, and methods, alongside documentation requirements to ensure accountability. Embedding such provisions in Articles 14 and 26 of the AI Act would support a proportionate and effective oversight regime.

In contract law, by contrast, the focus is not on third-party risks (to society or unrelated individuals) but on enabling effective and efficient transactions, and mitigating counterparty risk. While EU law has long grappled with, and developed solutions for, the challenges of automated contracting, autonomy of agents gives rise to specific concerns surrounding errors and ultra vires acts. Building on recent proposals by fellow academics, we make specific suggestions for ensuring the validity of agent-mediated contracts, protections for principals against agents acting outside of their safety perimeter, and for a traffic light system of staggered authorizations that may guide future machine to machine contractual interactions.

The following table summarizes our policy recommendations concerning the AI Act (Proposals #1-3) and contract law (Proposals #4-6).

| No. | Proposal | Domain | Key Content | Legal Basis/ Internal References |
|---|---|---|---|---|
| 1 | Mandatory Oversight Protocol for Autonomous Agents | AI Act – High-Risk Rules (Monitoring) | • Specific rules/guidance supplementing Art. 26(5), (6) for agentic autonomy<br>• Minimum review intervals; defined scope of oversight activities<br>• Structured documentation mandate for oversight outcomes | *Art. 26(5), (6) AI Act; Sections 3.3.3.2 and 3.6. of the paper* |
| 2 | Deployer-to-Provider Reclassification under Annex I A | AI Act – Value Chain Rules (Classification) | • Art. 25(1)(c) applicable even if original provider's intended purpose is sole criterion under Annex I A legislation<br>• Deployer repurposing non-high-risk system (e.g., agentic GPAI) for high-risk use → reclassification as provider<br>• Triggers contextual high-risk obligations (risk management, performance, cybersecurity, robustness) | *Art. 25(1)(c) AI Act; Annex I A; Sections 3.4.3 and 3.6. of the paper* |
| 3 | Extension of Collaboration Duties to GPAI Model Providers | AI Act – Value Chain Rules (Collaboration) | • Art. 25(2) collaboration duties extended to GPAI model providers<br>• Right of downstream deployers-turned-providers to compliance-relevant information<br>• Covers, e.g., data governance obligations (Art. 10 AI Act) | *Art. 25(2) AI Act; Sections 3.4.5 and 3.6. of the paper* |
| 4 | General Permission and Non-Discrimination for Autonomous Contract Conclusion | Agentic Contracting (Market-Enabling Law) | • Contract validity independent of natural-person review of agent action<br>• Default attribution to deploying principal; exceptions for bad faith / evident excess of authority<br>• Consistent with Electronic Communications Convention and UNCITRAL Model Law | *Ecommerce Directive and Member State contract law; Sections 4.1 and 4.3 of the paper* |

| 5 | Statutory List of Non-Delegable / Presumptively Ultra Vires Transactions | Agentic Contracting (Market-Enabling Law) | • Defined transaction types excluded from or restricted for agent conclusion<br>• Categories: core-asset dispositions, structural corporate acts, value thresholds (e.g., > € 1 million)<br>• Protection against significantly adverse commitments from system error / manipulation (e.g., prompt injection) | *Ecommerce Directive and Member State contract law; Sections 4.2 and 4.3 of the paper* |
| --- | --- | --- | --- | --- |
| 6 | "Authority Traffic Light" with Machine-Readable Attestations | Agentic Contracting (Market-Enabling Law) | • Externally unrestrictable authority; internal limits not binding on third parties absent knowledge of excess<br>• Standardized authority grades: colour code ("traffic light") for humans; metadata/attribute certificates for machines<br>• Legally relevant signal shifted from conversational layer to machine-readable/-verifiable authorization statements for M2M communications | *Ecommerce Directive and Member State contract law; Sections 4.2 and 4.3 of the paper* |

**Table 3: Summary of policy recommendations**

Going forward, developing technical standards for validating agentic systems is also critical. These standards should not fall solely to GPAI providers but emerge from collaboration among regulators, industry, and academia. Although beyond the scope of this paper, standardised testing, evaluation, and certification frameworks will be necessary to ensure that autonomous agents act safely, reliably, and transparently within their intended domains. Here we need clear principles to ensure the validity of the assessment, in terms of model, context and commercial independence.

In sum, agentic AI marks a significant shift in autonomous reasoning and execution across a wide range of contexts. Effective governance requires recognising their distinct characteristics: contextuality, autonomy, and heterogeneity. Regulating agents as context-specific systems will better protect the public while enabling responsible innovation. Because agentic systems are already being deployed, timely regulatory action is essential. It is for this reason that we favour amending existing regulations and legal provisions, over proposing an entirely new suite of regulation as was the case for GPAI systems. We do so, however, acknowledging that agentic AI's autonomy and scale introduce new forms of risk that ultimately may extend beyond current legal frameworks. Current AI agents are largely text-based, digital agents embedded in workflows within organisations. Autonomous AI systems embedded in physical agents, like co-bots, will pose further regulatory challenges to ensure safe human-AI interaction within a shared physical space (like road traffic or a factory workplace) that are of an entirely different nature. Ideally, legal frameworks would be as flexible and adaptable as the technology itself; in practice, this will likely require continued, detailed updates to sectoral legislation and possibly new horizontal regulatory layers.